\documentclass[twocolumn, pra, showpacs,superscriptaddress]{revtex4}
\usepackage{graphicx}
\usepackage{amsmath}
\usepackage{bm}
\usepackage{float}
\usepackage{color}
\usepackage{sidecap}
\usepackage{amsfonts}

\begin{document}
\title{Symmetry classification of spin-orbit coupled spinor Bose-Einstein condensates}
\author{Z. F. Xu}
\affiliation{Department of Physics, University of Tokyo, 7-3-1 Hongo, Bunkyo-ku, Tokyo 113-0033, Japan}
\author{Y. Kawaguchi}
\affiliation{Department of Physics, University of Tokyo, 7-3-1 Hongo, Bunkyo-ku, Tokyo 113-0033, Japan}
\author{L. You}
\affiliation{State Key Laboratory of Low Dimensional Quantum Physics,
Department of Physics, Tsinghua University, Beijing 100084, China}
\author{M. Ueda}
\affiliation{Department of Physics, University of Tokyo, 7-3-1 Hongo, Bunkyo-ku, Tokyo 113-0033, Japan}
\affiliation{Macroscopic Quantum Control Project, ERATO, JST, Bunkyo-ku, Tokyo 113-0033, Japan}

\date{\today}

\begin{abstract}
  We develop a symmetry classification scheme to find ground states of
pseudo spin-1/2, spin-1, and spin-2 spin-orbit coupled spinor Bose-Einstein condensates,
and show that as the SO(2) symmetry of simultaneous spin and space rotations is broken into discrete cyclic groups,
various types of lattice structures emerge in the absence of a lattice potential, examples include 
two different kagaome lattices for pseudo spin-1/2 condensates
and a nematic vortex lattice in which uniaxial and biaxial spin textures align alternatively for spin-2
condensates. For the pseudo spin-1/2 system, although mean-field states 
always break time-reversal symmetry, there exists a time-reversal invariant many-body
ground state, which is fragmented and expected to be observed in a micro-condensate.

\end{abstract}

%03.75.Mn, 03.75.Hh, 67.85.Fg}
%\pacs{03.75.Mn, 67.85.Fg, 67.85.Jk}
%03.75.Hh   Static properties of condensates; thermodynamical, statistical, and structural properties
%67.85.Jk   Other Bose-Einstein condensation phenomena
%03.75.Kk, 03.75.Mn, 67.85.Fg
%05.30.Jp   Boson systems (for static and dynamic properties
%of Bose-Einstein condensates, see 03.75.Hh and 03.75.Kk; see also 67.10.Ba Boson degeneracy in quantum fluids)
%67.85.Fg   Multicomponent condensates; spinor condensates
%05.30.Jp   Boson systems
%67.85.-d   Ultracold gases, trapped gases (see also 03.75.-b Matter waves in quantum mechanics)
%03.75.Mn   Multicomponent condensates; spinor condensates
%03.75.Kk   Dynamic properties of condensates, collective and hydrodynamic excitations, superfluid flow
%67.60.Bc   Boson mixtures
%67.86.De   Dynamic properties of condensates, excitations, and superfluid flow
%05.45.Xt   Synchronization; coupled oscillators
%05.45.Pq   Numerical simulations of chaotic systems

\maketitle

\section{Introduction}
Ultracold quantum gases have provided an exceptionally idealized
playground for emulating condensed matter systems \cite{bloch2008}.
Recently, Lin {\it et al.} have achieved a breakthrough in
quantum simulations of condensed matter systems by realizing an
effective spin-orbit (SO) coupling in $^{87}$Rb atoms
\cite{lin2011}. The SO coupling produces non-abelian gauge fields
\cite{dalibard2011}, plays a key role in spintronics
\cite{zutic2004} such as spin-polarized transport, spin injection,
and spin relaxation, and yields many other interesting phenomena such as
the quantum spin Hall effect and topological insulators
\cite{qi2010}. The experimental realization of synthetic gauge
fields has stimulated tremendous efforts on SO-coupled
quantum gases
\cite{stanescu2008,wu2011,wang2010,ho2011,yip2011,xu2011,
kawakami2011,xqxu2011,hu2012,sinha2011,radic2011,zhou2011,deng2011}.

The SO interaction couples spin and linear momentum,
thereby significantly modifying single-particle spectra.
For the Rashba-type SO coupling \cite{ruseckas2005,campbell2011,sau2011,xu2011b}, the conserved quantity
is $L_z+F_z$, where $L_z$ and $F_z$ are the projected
orbital and spin angular momenta along the $z$-axis, respectively.
In homogeneous systems, an axisymmetric SO coupling will change the parabolic
single-particle spectrum of a spin-$F$ atom into $2F+1$ energy bands with the lowest one featuring a Mexican
hat, causing a circular degeneracy of single-particle
ground states. For trapped systems, such circular dengeneracy is
reduced into double and no degeneracy for half-integer and integer spin systems,
respectively.

In spinor Bose-Einstein condensates (BECs) \cite{ueda2010}, the coupling between spin
and linear momentum will cooperate or compete with spin-dependent or
spin-independent interactions, giving rise to
many exotic ground states with or without harmonic trapping potentials,
such as plane-wave, stripe, triangular, and square lattice phases \cite{wang2010,xu2011,kawakami2011}.
Firstly, the plane-wave phase and the stripe phase are found in 
SO-coupled pseudo spin-1/2 or spin-1 BECs \cite{wang2010}.
Later, we found two different lattice phases where each spin component shows
trianguar- or square-lattice density distributions in SO-coupled
spin-2 BECs with cyclic interactions \cite{xu2011}.
As an axisymmetry harmonic trap is turned on, more phases are
found as ground states of SO-coupled spin-1/2 BECs \cite{hu2012,sinha2011}.
Although many phases are found in SO-coupled spinor BECs,
a systematic understanding on them is still elusive as previous results
largely depend on numerical solutions of the coupled Gross-Pitaevskii equations.
In this article, alternatively we present a symmetry classification scheme to 
investigate ground states of SO-coupled spinor condensates.
We can then not only understand different lattice phases already found,
but also find two different kagaome-lattice phases and a nematic vortex lattice phase,
both of which emerge spontaneously without lattice potentials.

This paper is organized as follows. Section II describes the model Hamiltonian 
used in the present paper. Section III analyzes various phases that spontaneously 
emerge in SO-coupled spinor condensates based on symmetry considerations. 
The following three sections discuss properties of the three typical phases: 
triangular-, square-, and kagome-lattice phases. Section VII discusses a 
possibility of a fragmented ground state of a SO-coupled spin-1/2 system 
with time-reversal symmetry. Section VIII summarizes the main results of this paper.

\section{Model Hamiltonian} 
\label{model}
We consider a SO-coupled spinor BEC
with $N$ atoms, including pseudo spin-1/2, spin-1, and spin-2 condensates in a pancake-shaped
quasi-two-dimensional harmonic potential.
The effective Hamiltonian is given by $\mathcal{H}=\mathcal{H}_0+\mathcal{H}_{\rm int}$, with 
\begin{eqnarray}
  \mathcal{H}_0=\int d\bm{\rho}\hat{\psi}^{\dag}\left[\frac{\mathbf{p}^2}{2M}+\mathcal{V}_o
    +\frac{v}{F}(p_xF_x+p_yF_y)\right]\hat{\psi},
  \label{singleparticle}
\end{eqnarray}
where $\hat{\psi}=(\hat{\psi}_{F},\dots,\hat{\psi}_{-F})^T$, 
$M$ is the atomic mass, $\bm{\rho}\equiv(x,y)$, $\mathcal{V}_o=M\omega_{\perp}^2\bm{\rho}^2/2$,
$v (>0)$ describes the strength of the SO coupling,
and $F_{x,y}$ are the spin-$F$ matrices.

For the $F=1/2$ case, we have
\begin{eqnarray}
  \mathcal{H}_{\rm int}&=&\frac{1}{2}\int d\bm{\rho}\left( g\hat{n}_{1/2}^2+g\hat{n}_{-1/2}^2+2g'\hat{n}_{1/2}\hat{n}_{-1/2}\right)
  \nonumber\\
  &=&\frac{1}{2}\int d\bm{\rho}\left(\alpha \hat{n}^2+\beta\hat{S}_z^2\right),
  \label{spinhalf}
\end{eqnarray}
where $g$ and $g'$ denote the strengths of the intra- and inter-component contact interactions, respectively.
Here, $\hat{n}_{\pm1/2}=\hat{\psi}_{\pm1/2}^{\dag}\hat{\psi}_{\pm1/2}^{\dag}\hat{\psi}_{\pm1/2}\hat{\psi}_{\pm1/2}$,
$\hat{n}=\hat{n}_{1/2}+\hat{n}_{-1/2}$, $\hat{S}_z=\hat{n}_{1/2}-\hat{n}_{-1/2}$, $\alpha=(g+g')/2$,
and $\beta=(g-g')/2$.

For the integer spin cases, we take the same interaction Hamiltonian
from spinor BECs \cite{ueda2010}.  
For the $F=1$ case, the interaction part is given by 
\begin{eqnarray}
  \mathcal{H}_{\rm int}=\frac{1}{2}\int d\bm{\rho}\left(\alpha \hat{\psi}_i^{\dag}\hat{\psi}_j^{\dag}\hat{\psi}_j\hat{\psi}_i
  +\beta\hat{\psi}_i^{\dag}\hat{\psi}_k^{\dag}\vec{F}_{ij}\cdot\vec{F}_{kl}\hat{\psi}_l\hat{\psi}_j\right),
  \label{spinone}
\end{eqnarray}
where $\alpha$ and $\beta$ give the strengths of density-density and spin-exchange interactions, respectively.
Here, the indices that appear twice are to be summed over $-F, \dots, F$.
For the $F=2$ case, the interaction Hamiltonian is given by
\begin{eqnarray}
  \mathcal{H}_{\rm int}&=&\frac{1}{2}\int d\bm{\rho}\left(\alpha \hat{\psi}_i^{\dag}\hat{\psi}_j^{\dag}\hat{\psi}_j\hat{\psi}_i
  +\beta\hat{\psi}_i^{\dag}\hat{\psi}_k^{\dag}\vec{F}_{ij}\cdot\vec{F}_{kl}\hat{\psi}_l\hat{\psi}_j\right.
  \nonumber\\
  &&+\left.\gamma (-1)^{i+j}\hat{\psi}_i^{\dag}\hat{\psi}^{\dag}_{-i} \hat{\psi}_j\hat{\psi}_{-j}\right),
  \label{spintwo}
\end{eqnarray}
where $\alpha$, $\beta$ and $\gamma$ give the strengths of the density-density, spin-exchange,
and singlet-pairing interactions, respectively.
Defining the harmonic-oscillator length $a_{\perp}\equiv\sqrt{\hbar/M\omega_{\perp}}$, we introduce
dimensionless parameters $v'=v/\omega_{\perp}a_{\perp}$ and
$(\alpha',\beta',\gamma')=(\alpha,\beta,\gamma)N/\hbar\omega_{\perp}$.

\section{Symmetry analysis} 
\label{symmetry}

Under a mean-field approximation, we assume
that all atoms are condensed into a common single-particle state,
and therefore the symmetries of the Hamiltonian are spontaneously
broken. We can classify the ground states according to the remaining symmetries \cite{volovik1985,bruder1986,makela2007,yip2007,kawaguchi2011}.

\begin{figure*}[t]
\centering
\includegraphics[width=4.1in]{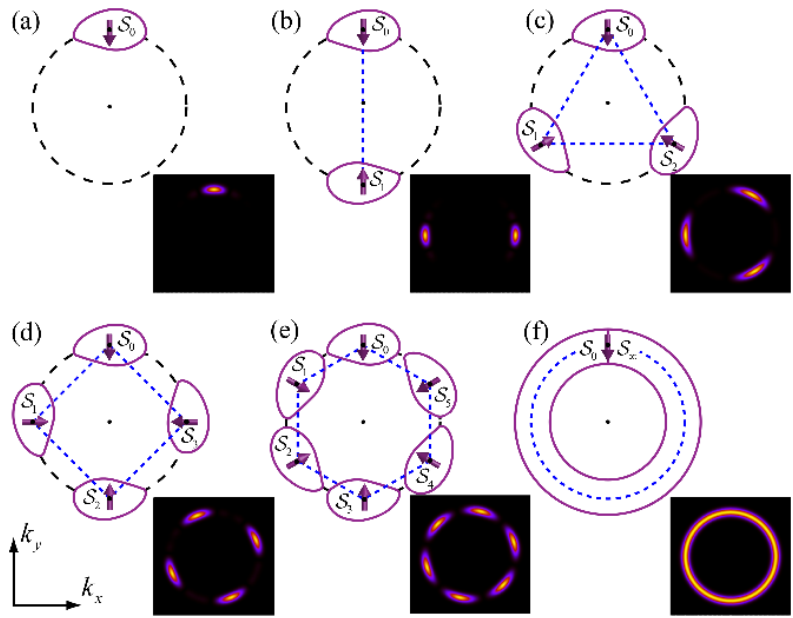}
\caption{(Color online). Schematic diagrams for states preserving the
combined symmetry of spin-space rotation $\mathcal{C}_{nz}$,
gauge transformation, and time reversal. 
Figures (a)--(f) correspond to $n=1$, $2$, $3$, $4$, $6$, and $\infty$,
respectively, where arrows denote the direction of spin, solid closed loops indicate
regions with nonzero momentum distributions $|\psi(k)|^2$, and
$n$ points on the circle are denoted as $\mathcal{S}_0$, \dots, $\mathcal{S}_{n-1}$,
respectively.
Characteristic momentum distributions 
of the ground state $\psi$ for a trapped system
are shown in the lower right corner of each figure.
Black (yellow) color refers to the region of small (large) amplitude. }
\label{fig1}
\end{figure*}

For spinor BECs \cite{ueda2010} without SO-coupling, we usually focus only on
the spin degrees of freedom because they are decoupled from the orbital degrees of freedom.
As a result, the Hamiltonian is invariant under the U(1) global gauge transformation, 
the SO(2) spin rotation along $z$-axis for the $F=1/2$ case or
the SO(3) spin rotation for the integer-spin cases, and the time reversal $\mathcal{T}\equiv e^{-i\pi F_y}\mathcal{K}$, 
where $\mathcal{K}$ takes complex conjugation.
In contrast, for SO-coupled spinor BECs, to make the Hamiltonian invariant, we should simultaneously rotate the spin and the space.
Therefore, the Hamiltonian $\mathcal{H}$ described in Sec.\ref{model} is invariant under the global
U(1) gauge transformation, simultaneous SO(2) global spin and space (spin-space) rotations, and time reversal $\mathcal{T}$.

According to the symmetry classification scheme \cite{volovik1985,bruder1986,makela2007,yip2007,kawaguchi2011},
our task is: (1) to find the full symmetry group of the Hamiltonian, which in the present case is $\rm G=U(1)\times SO(2)\times \mathcal{T}$;
(2) to list all subgroups H of G; (3) to find the order parameter that is invariant under H.
If the order parameter can be uniquely determined from $h\psi=\psi$ for $\forall h\in \rm H$,
the state $\psi$ is an inert state, which is always a stationary point of the energy functional. 
On the other hand, if the order parameter that is invariant under H is not uniquely determined, 
we need to minimize the energy functional within the restricted manifold. 
Such a state is called a non-inert state \cite{bruder1986,makela2007,yip2007,kawaguchi2011}.
The high-symmetry states in the SO-coupled system are all non-inert states.
To see this more clearly, we consider the eigenstates of simultaneous spin-space rotation $\mathcal{C}_{nz}$,
where $\mathcal{C}_{nz}\equiv\mathcal{R}_F (2\pi/n)\mathcal{R}_{\rho}(2\pi/n)$ 
is the generator of a discrete cyclic subgroup of SO(2) with $\mathcal{R}_F=\exp(-iF_z2\pi/n)$ 
and $\mathcal{R}_{\rho}$ being respectively the $2\pi/n$ spin and space rotation operators about the $z$-axis. 
For the integer-spin cases, $\mathcal{C}_{nz}$ has $n$ different eigenvalues $\exp(-i2\mathbb{N}\pi/n)$ with $\mathbb{N}=0,\dots,n-1$,
whereas for the $F=1/2$ case, $\mathcal{C}_{nz}$ has $n$ different eigenvalues $\exp\{-i(2\mathbb{N}+1)\pi/n\}$
because $(\mathcal{C}_{nz})^n=-1$.
If the eigenvalue of $\mathcal{C}_{nz}$ is not $1$,
we can infer that this state is invariant under the combined U(1) gauge transformation
and spin-space rotation $\mathcal{C}_{nz}$, namely, 
$\mathcal{O}_{nz}(\mathbb{N})\psi = \psi$ 
where $\mathcal{O}_{nz}(\mathbb{N})=\exp\{i(2\mathbb{N}+1)\pi/n\}\mathcal{C}_{nz}$ for the $F=1/2$ case,
and $\mathcal{O}_{nz}(\mathbb{N})=\exp(i2\mathbb{N}\pi/n)\mathcal{C}_{nz}$ for the $F=1$ and $2$ cases.
For a spin-$F$ system, we can construct eigenstates of $\mathcal{O}_{nz}(\mathbb{N})$ by using a complete basis set of plane waves as
\begin{eqnarray}
  \psi=\sum\limits_{j=0}^{n-1}\left[\mathcal{O}_{nz}(\mathbb{N})\right]^j\sum\limits_{\mathbf{k}\in \bar{\mathcal{S}},\sigma}
  D^{n}_{\sigma}(\mathbf{k})e^{i\mathbf{k}\cdot\bm{\rho}}|\sigma\rangle,
  \label{rotation}
\end{eqnarray}
where $\mathbf{k}$ is the wave vector,
$\bar{\mathcal{S}}=\{\mathbf{k}| -\pi/n+\bar{\varphi}\le\varphi_k<\pi/n+\bar{\varphi}, \varphi_k\equiv\arg(k_x+ik_y)\}$
with $\bar{\varphi}$ being arbitrary, $|\sigma\rangle$ ($\sigma=-F,-F+1,\dots,F$)
denotes the spin state with $M_F=\sigma$ under the $z$-axis quantization,
and $D_{\sigma}^{n}(\mathbf{k})$ are the expansion coefficients.
Unless we know the detail of $D_{\sigma}^{n}(\mathbf{k})$, 
we cannot uniquely determine the state that is invaraint under ${\rm H}=\{\mathbb{E}, \mathcal{O}_{nz}(\mathbb{N}),\dots,\mathcal{O}_{nz}^{n-1}(\mathbb{N})\}$
with $\mathbb{E}$ being the identity operator.
Similar arguments can be applied to other subgroups of G.

Since there are no inert states for spin-orbit coupled spinor BECs,
we cannot find the order parameter by analyzing its symmetry only.
However, such a difficulty can be alleviated if we take the $\it ansatz$ that  the state
is a superposition of several degenerate single-particle ground states of the Hamiltonian in Eq. (\ref{singleparticle})
with $\omega_{\perp}=0$:
\begin{eqnarray}
  \psi=\sum\limits_{j=0}^{n-1}e^{i\phi_j}\mathcal{PW}\left(k=k_g,\varphi_k=\bar{\varphi}+j\frac{2\pi}{n}\right),
  \label{orderparameters}
\end{eqnarray}
where $\mathcal{PW}(k_g,\varphi_{k})=e^{i\mathbf{k}_g\cdot\bm{\rho}}\zeta_{-F}(\varphi_{k})$
are degenerate single-particle ground states
with $k_g=mv/\hbar$, $\varphi_k=\arg(k_x+ik_y)$,
and $\bar{\varphi}$ being arbitrary. For the cases of $F=1/2$, $1$ and $2$,
we have respectively
\begin{eqnarray}
  &\zeta_{-1/2}(\varphi_k)=(1,-e^{i\varphi_k})^T/\sqrt{2},&\nonumber\\
  &\zeta_{-1}(\varphi_k)=(e^{-i\varphi_k},-\sqrt{2},e^{i\varphi_k})^T/2,&\nonumber\\
  &\zeta_{-2}(\varphi_k)=(e^{-2i\varphi_k},-2e^{-i\varphi_k},\sqrt{6},-2e^{i\varphi_k},
  e^{2i\varphi_k})^T/4.&
  \label{planewaves}
\end{eqnarray}
The corresponding spin expectation value is  anti-parallel to the wave vector: 
$(\langle F_x\rangle, \langle F_y \rangle)=-F(\cos\varphi_k,\sin\varphi_k)$.
If we take $\phi_j=j2\mathbb{N}\pi/n$, the state $\psi$ in Eq. (\ref{orderparameters})
is invariant under $\mathcal{O}_{nz}(\mathbb{N})$. This state is a special case of 
that in Eq. (\ref{rotation}) with the set $\bar{\mathcal{S}}$ shrinks into only one 
element and the allowed values of $\mathbf{k}$ are fixed at discrete points
as $D_\sigma^n({\bf k})=\delta(k-k_g)[\zeta_{-F}(\bar{\varphi})]_\sigma$.
Figure 1 illustrates schematic diagrams for the states of Eq. (\ref{orderparameters}),
where $n$ single-particle ground states $\mathcal{PW}(k_g,\varphi_{k})$ are denoted by
points $\mathcal{S}_j$.

\begin{figure*}[t]
\centering
\includegraphics[width=4.1in]{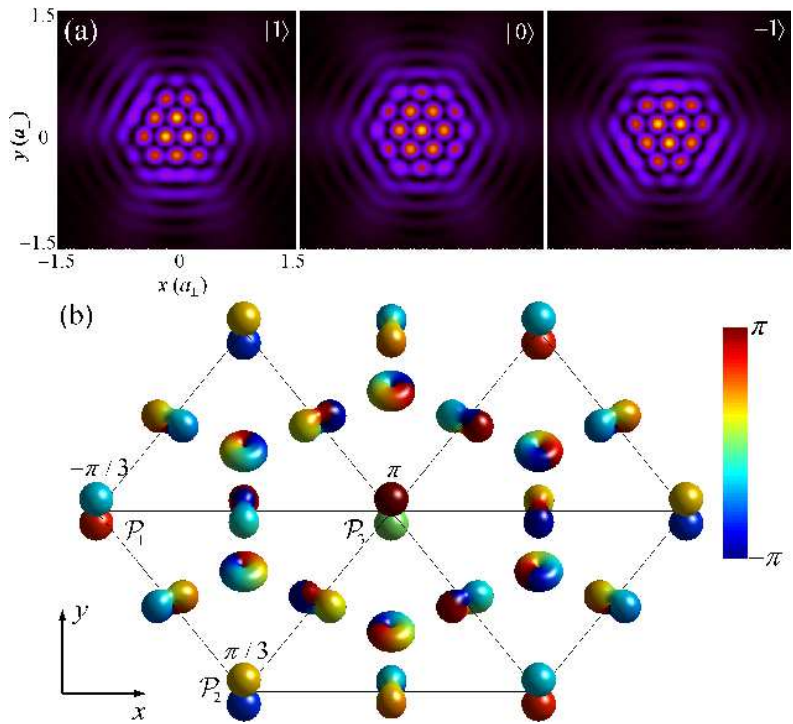}
\caption{(Color online). Ground states of trapped SO-coupled spin-1 condensates,
with $v'=15$, $\alpha'=0.5$ and $\beta'=0.1$.
(a) Density distributions of three spin components $M_F=1,0,-1$ from left to right.
(b) Spatial variation of the corresponding order parameter visualized by plotting
$\sum_{m}\xi_{m}Y_F^{m}(\theta,\varphi)$,
where $Y_F^{M_F}$ is the spherical harmonics and $\xi_m$ is the $m$-th component
of the spin wave function. }
\label{fig2}
\end{figure*}

We choose the $\it ansatz$ in Eq. (\ref{orderparameters})
based on previous understanding on a 2D system and our numerical results 
given in the following sections.
For a 2D homogenous system with $\omega_{\perp}=0$,
as discussed in Refs. \cite{wang2010,xu2011},
plane-wave and stripe phases are found to be the ground states for
the $F=1/2$, $1$ and $2$ condensates \cite{wang2010,xu2011},
whereas triangular and square lattice phases exist
only for spin-2 condensates with cyclic interactions \cite{xu2011}.
Such phases can all be described by Eq. (\ref{orderparameters}).
For a trapped system, translation symmetry is broken, and linear momentum is no
longer conserved.  Therefore, in momentum space, there are infinite points forming
regions, where momentum distributions $|\psi(\mathbf{k})|^2$ of the
ground state $\psi$ are nonzero. In Fig. \ref{fig1}, we use solid closed
loops (circles for $n=\infty$ ) to denote such regions.
In the case of strong SO couplings, we can approximate low-lying single-particle eigenstates as \cite{jacob2009,sinha2011}
\begin{eqnarray}
  \Psi_{\bar{n},m}(k')\propto k'^{-1/2}e^{-(k'-v')^2/2}H_{\bar{n}}(k'-v')e^{im\varphi_k}\zeta_{-F}(\varphi_k),
  \label{eigenstates}
\end{eqnarray}
with eigenenergies
$E_{\bar{n},m}=(m+1/2)^2/2v'^2+\bar{n}+1/2-v'^2/2$, $(m^2+1/4)/2v'^2+\bar{n}+1/2-v'^2/2$,
and $(m^2+3/4)/2v'^2+\bar{n}+1/2-v'^2/2$ for $F=1/2$, $1$ and $2$ respectively.
Here, $\bar{n}=0,1,2,\dots$, $m=0,\pm1,\pm2, \dots$, $k'=ka_{\perp}$, and $H_{\bar{n}}$ are Hermite polynomials.
When the interatomic interactions are weak, the ground states can be constructed only from
the states with $\bar{n}=0$, which are superpositions of single-particle
eigenstates $\zeta_{-F}(\mathbf{k})$ with weight  $k'^{-1/2}e^{-(k'-v')^2/2}\simeq v'^{-1/2}e^{-(k'-v')^2/2}$.
In the limit of $\omega_{\perp}\rightarrow0$, $e^{-(k'-v')^2/2}\rightarrow \delta(k'-v')/\sqrt{2\pi}$,
which implies that the ground states will involve momenta with its magnitude close to $k_g$.
Furthermore, by choosing specific weak interactions, axial rotational symmetry in the single-particle 
states of Eq. (\ref{eigenstates}) can break into discrete rotation symmetry,
and 2D lattice phases appear, as already found in the $F=1/2$ case \cite{hu2012,sinha2011}. 
In the limit of weak trapping potential, strong SO couplings, and weak interatomic interactions,
there are ground states whose momentum distributions split into several small regions 
as those illustrated in Fig. \ref{fig1}. 
As long as the area of the allowed momentum is small enough, 
these ground states can still be well described by Eq. (\ref{orderparameters}).

The validity of the {\it ansatz} has been numerically confirmed \cite{numerical},
because we find several ground states which can be understood by Eq. (\ref{orderparameters}).
In the lower right corner of each figure in Fig. \ref{fig1},  we
illustrate the characteristic momentum distributions of these ground states.
For the case of $n=1$ and $2$, the state in Eq. (\ref{orderparameters})
can describe respectively the plane-wave phase and the stripe phase found in a homogeneous system \cite{wang2010,xu2011}.
For the case of $n=\infty$, the state is invariant under the combined SO(2) spin-space rotations and U(1) gauge transformation,
and has been predicted for the $F=1/2$ case \cite{hu2012,sinha2011},
where single-particle eigenstates of Eq.~(\ref{eigenstates}) will be stabilized in some parameter regions.
For the $F=1$ and $F=2$ cases, similar arguments can be applied, and more single-particle eigenstates
of Eq.~(\ref{eigenstates}) can be stabilized.

In the following, we start from the ansatz of Eq. (\ref{orderparameters}). 
and find several high-symmetry states, where a set of $\phi_j$'s
are determined by requiring the wave function of Eq. (\ref{orderparameters}) to preserve a certain symmetry. 
Some of them  become ground states of a trapped SO-coupled system, 
where triangular, square and kagaome lattices are formed
spontaneously without lattice potentials. 
These states can be described by Eq. (\ref{orderparameters})
with $n=3$, $4$, and $6$, respectively.

Before discussing lattice phases, we point out that
(1) the spin and space rotation operations commute with time reversal
$[\mathcal{T},\mathcal{C}_{nz}]=0$, and that 
(2) for the $F=1/2$ case, the ground-state order parameter $\psi$ always breaks time-reversal symmetry
because $\mathcal{T}^2=-1$, which means $\psi$ will only be invariant under $\mathcal{C}_{nz}$
or $\mathcal{TC}_{nz}$, and possibly combined with the U(1) gauge transformation for each $n$.
For the integer-spin case, the ground-state order parameter will be invariant under one or all of the three operations:
$\mathcal{T}$, $\mathcal{C}_{nz}$ and $\mathcal{TC}_{nz}$,
and possibly combined with the U(1) gauge transformation for each $n$.

\section{Triangular-lattice phase} 
A triangular-lattice phase is described by
Eq.~(\ref{orderparameters}) with $n=3$ as
\begin{eqnarray}
  \psi&=&e^{i\phi_0}e^{ik_gx}\zeta_{-F}(0)+e^{i\phi_1}e^{ik_g(-x/2+\sqrt{3}y/2)}\zeta_{-F}(2\pi/3)\nonumber\\
  &&+e^{i\phi_2}e^{ik_g(-x/2-\sqrt{3}y/2)}\zeta_{-F}(4\pi/3),
  \label{triangular}
\end{eqnarray}
where we take $\bar{\varphi}=0$ as an example.
According to the symmetry classification scheme described in Sec.\ref{symmetry},
we can apply specific symmetries on the state of Eq. (\ref{triangular})
to evaluate the value of $\phi_j$.
We find the only possible symmetry is described by the group 
${\rm H}=\{\mathbb{E}, \mathcal{O}_{3z}(\mathbb{N}),\mathcal{O}_{3z}^{2}(\mathbb{N})\}$,
resulting in $\phi_j=j2\mathbb{N}\pi/3+\rm const$.
Meanwhile, the states of Eq. (\ref{triangular}) with different values of $\phi_j$ 
are connected by a global U(1) phase change and a global coordinate translation.
This is because for arbitrary change $\delta\phi_j$ in $\phi_j$,
there are always solutions for $(\delta x,\delta y)$ and $\delta U$ that satisfy
\begin{eqnarray}
  &\exp\{i[\delta \phi_0+k_g\delta x+\delta U)\}=1,&\nonumber\\
  &\exp\{i[\delta \phi_1+k_g(-\delta x/2+\sqrt{3}\delta y/2)+\delta U)\}=1,&\nonumber\\
  & \exp\{i[\delta \phi_2+k_g(-\delta x/2-\sqrt{3}\delta y/2)+\delta U)\}=1,&
\end{eqnarray}
where $(\delta x, \delta y)$ and $\delta U$ describe 
the amount of the coordinate translation and the U(1) phase change, respectively.
As a result, there is only one type of triangular-lattice phase.
By choosing a proper center of lattice, such a ground state
becomes invariant under $\mathcal{O}_{3z}$. Note that in a harmonic trap,
the $\mathcal{O}_{3z}$ symmetry axis sometimes deviates from the trap center to
gain the interaction energies.

In Fig.~\ref{fig2}, we illustrate
a triangular-lattice phase which appears in the ground state of SO coupled spin-1 condensates with
$v'=15$, $\alpha'=0.5$ and $\beta'=0.1$,
showing triangular-lattice density distributions for each spin component.
Figure \ref{fig2}(b) shows the spatial variation of the corresponding order parameter.
At three lattice sites, the order paramters
can be written as $\mathcal{P}_1: (0,e^{-i\pi/3},0)^T$, $\mathcal{P}_2: (0,e^{i\pi/3},0)^T$
and $\mathcal{P}_3: (0,e^{i\pi},0)^T$.
To go from $\mathcal{P}_1$ to $\mathcal{P}_2$,
a $\pi$ spin rotation along $\mathcal{P}_1\mathcal{P}_2$ and the $e^{-i\pi/3}$ gauge transformation
are needed. Similar transformations are needed to go from $\mathcal{P}_2$ to $\mathcal{P}_3$
or from $\mathcal{P}_3$ to $\mathcal{P}_1$.
Going along the loop of $\mathcal{P}_1\mathcal{P}_2\mathcal{P}_3\mathcal{P}_1$,
the order parameter undergoes a $\pi$ spin rotation along an axis on the $x$-$y$ plane and
the $-\pi$ gauge transformation, 
which, however, does not imply that the mass circulation here is fractional with $-1/2$ winding \cite{fractional}.
To check whether this vortex is fractional or not, we should calculate the 
circulation \cite{ueda2010} along the loop of $\mathcal{P}_1\mathcal{P}_2\mathcal{P}_3\mathcal{P}_1$.
Numerically we found this vortex is not fractional.

For the triangular-lattice phase, two (three) spin components of spin-1/2 (spin-1) condensates
tend to occupy different spaces due to the interferences of three plane waves.
For spin-2 condensates, five spin components are divided into
three classes: (1) $M_F=2,-1$; (2) $M_F=1,-2$; (3) $M_F=0$.
In the cases of (1) and (2), two different spin components
show the same wave function, which can be
understood from Eq.~(\ref{triangular}). These properties
are consistent with our previous work \cite{xu2011}.

\begin{figure*}
\centering
\includegraphics[width=4.1in]{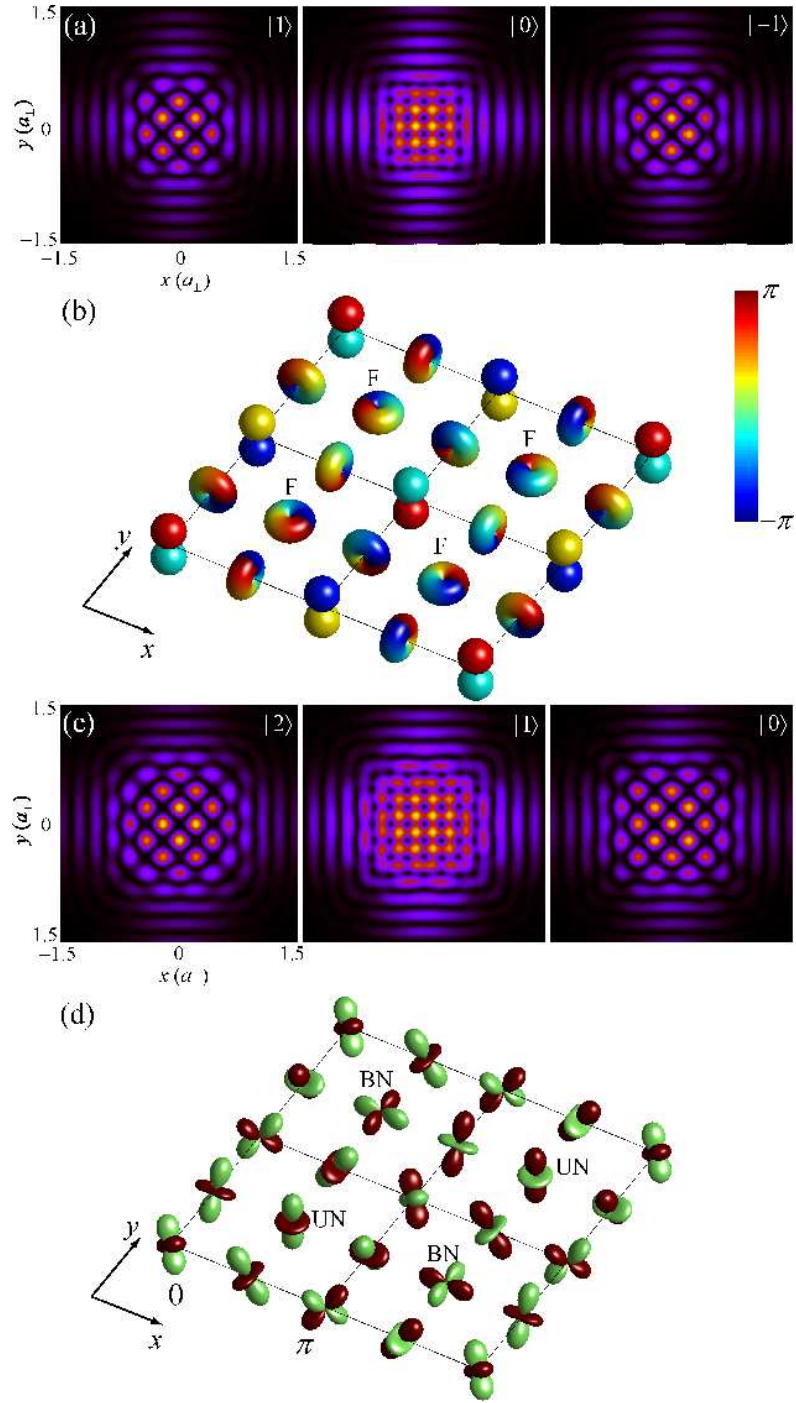}
\caption{(Color online). Ground states of trapped SO-coupled spinor BECs,
with $v'=15$. (a) Density distributions of $M_F=1$ (left),
$0$ (middle), $-1$ (right) of the spin-1 case with  $\alpha'=0.5$, $\beta'=-0.1$.
(b) The corresponding order parameter, where F is used to denote ferromagnetic vortex cores.
(c) Density distributions of $M_F=2$ (left), $1$ (middel), and $0$ (right) of the spin-2 case
with $\alpha'=0.5$, $\beta'=0.1$ and $\gamma'=-0.1$. (d) The corresponding order parameter.
The uniaxial and biaxial nematic vortex cores are denoted by UN and BN, respectivley.
Here, black (yellow) color refers to the low (high) density region.}
\label{fig3}
\end{figure*}

\begin{figure*}
\centering
\includegraphics[width=4.1in]{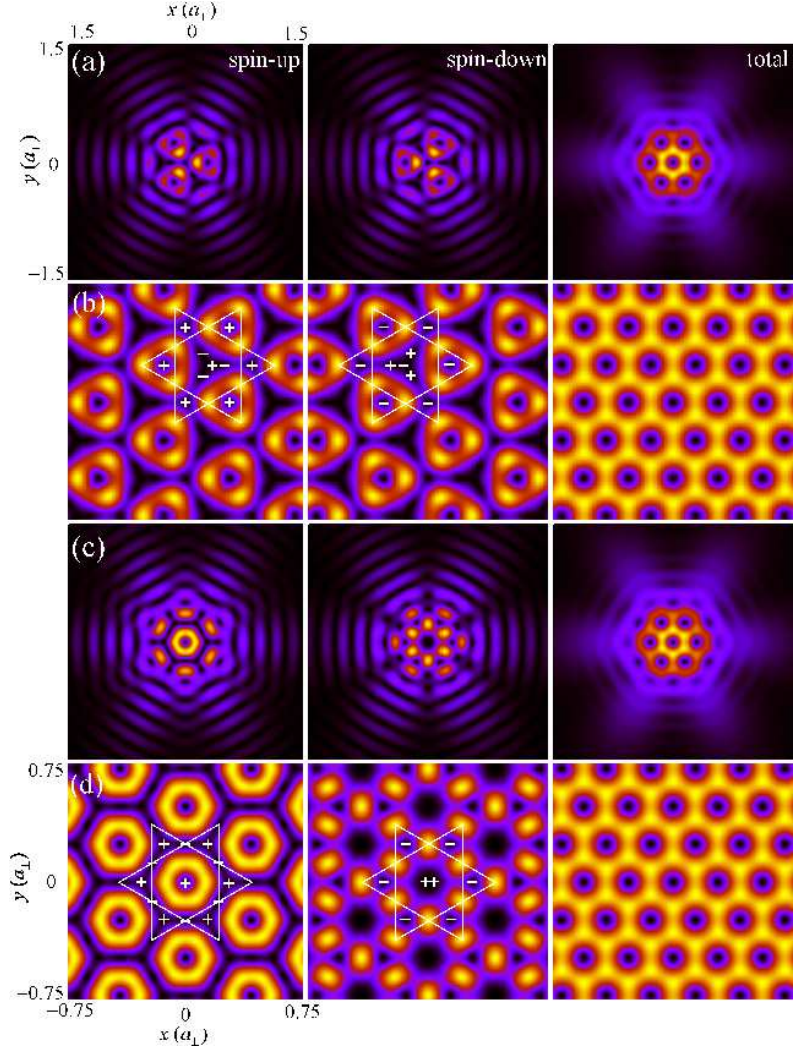}
\caption{(Color online). Ground states of trapped SO-coupled pseudo spin-1/2 BECs,
with $v'=15$ and (a,b) $\alpha'=0.3$, $\beta'=0.06$,
(c,d) $\alpha'=0.4$, $\beta'=0.08$. Here, black (yellow) color refers to the low (high) density region.
From left to right, we show the spin-up, spin-down and total density distributions.
Figures (a) and (c) are numerically calculated for a trapped system,
while (b) and (d) are obtained by superposition
of six single-particle states $\mathcal{PW}(\varphi_k)$.
In (b) and (d), we use symbols ``-'',``+'',  and ``++'' to denote vorticities
$-\hbar$, $\hbar$ and $2\hbar$, respectively. The size of figure (b) [(c)]
is the same as that of (d) [(a)]. }
\label{fig4}
\end{figure*}

\section{Square-lattice phase}  
When $n=4$, the state in Eq. (\ref{orderparameters}) can be simplified as
\begin{eqnarray}
  \psi&=&e^{i\phi_0}e^{ik_gx}\zeta_{-F}(0)+e^{i\phi_1}e^{ik_gy}\zeta_{-F}(\pi/2)\nonumber\\
  &+&e^{i\phi_2}e^{-ik_gx}\zeta_{-F}(\pi)+e^{i\phi_3}e^{-ik_gy}\zeta_{-F}(3\pi/2),
  \label{square}
\end{eqnarray}
where we again take $\bar{\varphi}=0$ as an example.
This state can only potentially be invariant under $\mathcal{O}_{nz}$ with $n=1,2,4$.
If the state is only invariant under $\mathcal{O}_{1z}$ or $\mathcal{O}_{2z}$,
$\psi$ cannot be uniquely determined.
Therefore, we require the state preserving $\mathcal{O}_{4z}(\mathbb{N})$,
resulting in $\phi_j=j\mathbb{N}\pi/2+\rm const$.

Similar to the triangular-lattice phase, there are other states 
which are related to the states invariant under $\mathcal{O}_{4z}(\mathbb{N})$
by simply doing a global lattice shift. They may be classified into the same class,
as they have the same lattice structure. The criterion for each class is 
determined by the parameter $\Delta \equiv(\phi_1+\phi_{3})-(\phi_0+\phi_{2})$.
The reason is given as follow:
When doing a global lattice shift, the state $\psi$ in Eq. (\ref{square}) changes
with 
\begin{eqnarray}
  \phi_0+k_gx\rightarrow\phi_0+k_g(x+\delta x),\nonumber\\
  \phi_1+k_gy\rightarrow\phi_1+k_g(y+\delta y),\nonumber\\
  \phi_2-k_gx\rightarrow\phi_2-k_g(x+\delta x),\nonumber\\
  \phi_3-k_gy\rightarrow\phi_3-k_g(y+\delta y).
\end{eqnarray}
Absorbing $\delta x$ and $\delta y$ into $\phi_j$, 
we find that there are two invariants $\phi_0+\phi_2$ and $\phi_1+\phi_3$.
Their difference $\Delta$ is also invariant under the global U(1) gauge transformation.

Furthermore, we find that for the states invariant under $\mathcal{O}_{4z}(\mathbb{N})$, 
$\Delta=(2\mathbb{M}+1)\pi$ when $\mathbb{N}=1, 3$;  
$\Delta=2\mathbb{M}\pi$ when $\mathbb{N}=0, 2$, ($\mathbb{M}\in \mathbb{Z}$).
We then define the criterion for two different types of square-lattice phases with different
lattice structures as (1) $\Delta=(2\mathbb{M}+1)\pi$ and 
(2) $\Delta=2\mathbb{M}\pi$.
The states that fall into the same class are connected by a global lattice shift.
By doing a proper lattice shift, the symmetry of the state $\psi$ can be described by 
${\rm H}=\{\mathbb{E}, \mathcal{O}_{4z}(\mathbb{N}),\mathcal{O}_{4z}^{2}(\mathbb{N}),\mathcal{O}_{4z}^{3}(\mathbb{N})\}$
with $\mathbb{N}=1, 3$ for the first class and $\mathbb{N}=0, 2$ for the second class.
For the integer-spin cases, the state $\psi$ in the second class
is also invariant under time reversal $\mathcal{T}$ \cite{statement}
by choosing a proper lattice center.
Numerically, we find all such ground states to be integer-spin condensates.
Again, in a harmonic trap, the $\mathcal{O}_{4z}$ symmetry axis does not
always coincide with the trap center to minimize the interaction energy. 

The ground states for the case of $\Delta=(2\mathbb{M}+1)\pi$ are found to exist in spin-1 BECs with $\beta<0$,
and spin-2 BECs with $\beta>0$ and $\gamma>0$ \cite{xu2011}.
Figures \ref{fig3}(a) and (b) show the corresponding ground-state density distributions
and order parameters for the spin-1 BECs, with $v'=15$, $\alpha'=0.5$, and $\beta'=-0.1$.
The spin component $M_F=0$ forms a square lattice with lattice constant $\pi/k_g$,
while the spin components $M_F=1$ and $-1$ fill the center of
squares alternatively, both forming a square lattice with lattice constant $\sqrt{2}\pi/k_g$.
The numerically obtained total density distributions are smooth for such ground states,
consistent with the prediction from
the corresponding state $\psi$ in Eq. (\ref{orderparameters}).
Only the component with $M_F=0$ involves vortices, with vortex
cores filled by the $M_F=1$ and $M_F=-1$ components having vorticity $\hbar$ and $-\hbar$, respectively.
Thus, the spin vortex lattice is filled by ferromagnetic vortex cores which are polarized
in the $z$ and $-z$ directions alternatively.
For the $F=2$ case, such phase has already been predicted in our previous work \cite{xu2011}.
We have found that spin components $M_F=2,0,-2$ show the same
density distribution which is different from those of the $M_F=1$ and $-1$ components.
Again, there is no fractional or integer vortex \cite{fractional}.

Another type of square lattice phase with $\Delta=2\mathbb{M}\pi$ appears in the
trapped SO-coupled spin-2 condensates with antiferromagnetic spin-dependent interactions.
Figure \ref{fig3}(c) shows ground-state density distributions of spin components
$M_F=2,1,0$ from left to right, with $v'=15$, $\alpha'=0.5$, $\beta'=0.1$,
and $\gamma'=-0.1$. The spin components $M_F=-1$ and $-2$ show the same density distributions
as $M_F=1$ and $2$, respectively.
In this phase, we find that two physical quantities
$|\langle\vec{F}\rangle|\equiv |\sum_{ij}\psi^*_i\vec{F}_{ij}\psi_j|/\sum_i\psi^*_i\psi_i$ and
$|\langle\Theta\rangle|\equiv|\sum_{ij}(-1)^{i+j}\psi_i^*\psi_{-i}^*\psi_j\psi_{-j}|/\sum_i\psi_i^*\psi_i$
are uniform and are equal to 0 and 1, respectively.
Figure \ref{fig3}(d) illustrates the corresponding order parameter,
showing a spin vortex lattice with uniaxial nematic (UN) and biaxial nematic (BN) vortex cores
aligning alternatively.

\section{Kagome-lattice phase}
The {\it ansatz} in Eq. (\ref{orderparameters}) with $n=6$
can be written as
\begin{eqnarray}
  \psi&=&e^{i\phi_0}e^{ik_gx}\zeta_{-F}(0)+e^{i\phi_1}e^{ik_g(x/2+\sqrt{3}y/2)}\zeta_{-F}(\pi/3)\nonumber\\
  &+&e^{i\phi_2}e^{ik_g(-x/2+\sqrt{3}y/2)}\zeta_{-F}(2\pi/3)+e^{i\phi_3}e^{-ik_gx}\zeta_{-F}(\pi)\nonumber\\
  &+&e^{i\phi_4}e^{ik_g(-x/2-\sqrt{3}y/2)}\zeta_{-F}(4\pi/3)\nonumber\\
  &+&e^{i\phi_5}e^{ik_g(x/2-\sqrt{3}y/2)}\zeta_{-F}(5\pi/3),
  \label{kagaome}
\end{eqnarray}
where we take $\bar{\varphi}=0$ as an example.
This state can describe kagaome-lattice structures.
Similar to the case of $n=4$, 
we note that four independent phases
\begin{eqnarray}
  &\Delta_0=\phi_0+\phi_2+\phi_4,\quad \Delta_1=\phi_0+\phi_3,&\nonumber\\
  &\Delta_2=\phi_2+\phi_5,\quad \Delta_3=\phi_4+\phi_1,&
\end{eqnarray}
are invariant under a global shift of the lattice.
Besides the one used to describe the global phase change, there
will be three independent phases which determine the structure of the lattice.
Therefore, there are infinite types of lattice structures of the state in Eq. (\ref{kagaome}).

According to symmetry classification scheme, we need to apply specific symmetries
to determine the state $\psi$. Only if the symmetry is high enough, 
the state $\psi$ can be uniquely determined.
For the $F=1/2$ case, two high symmetry groups are
generated by operators $\mathcal{O}_{6z}$ and $\mathcal{TO}_{6z}$, respectively,
whereas for the integer-spin cases, high symmetry groups can be generated
by only $\mathcal{O}_{6z}$ or $\{\mathcal{O}_{6z}, \mathcal{T}\}$.

(1) If we require the state is invariant under $\mathcal{O}_{6z}(\mathbb{N})$,
the state $\psi$ is uniquely determined with $\phi_j=j\mathbb{N}\pi/3+\rm const$.
For the integer-spin cases, the state is further invariant under time reversal $\mathcal{T}$ \cite{statement}
if $\mathbb{N}=0, 3$.

(2) For the $F=1/2$ case, there are states that preserve the $\mathcal{TO}_{6z}$ symmetry and break the $\mathcal{O}_{6z}$ symmetry. 
To determine such states, we start from the states invariant under $\mathcal{O}_{3z}(\mathbb{N})$ with $\mathbb{N}=0, 1, 2$.
The value of $\phi_j$ is determined as
$\phi_j=j2\mathbb{N}\pi/6+\text{mod}(j,2)\phi+\rm const$, where $\phi$ is arbitrary.
Furthermore, by requiring that the state be invariant under $\mathcal{TC}_{2z}$ \cite{statement}, 
we obtain $\mathbb{N}=1$ and $\phi=\pi/2+\mathbb{M}\pi$.

We numerically find two distinct classes of kagaome-lattice phases in a spin-1/2 BEC with
$\beta'>0$, whose symmetries are described by the group H generated respectively by (1) $\exp(i2\pi/3)\mathcal{C}_{3z}$ and $\mathcal{TC}_{2z}$
(with $\phi_j= j\pi/3+\bmod(j,2)(1/2+\mathbb{M})\pi +\rm const$)
and (2) $\exp\{i(\mathbb{M}+1/3)\pi\}\mathcal{C}_{6z}$ (with $\phi_j=j(\mathbb{M}+1/3)\pi+\rm const$), where 
the ground states with different values of $\exp(i\mathbb{M}\pi)$ in the same class
are time reversal with each other.
Figure \ref{fig4} shows the spin-up, spin-down and total density distributions
for numerically obtained order parameters in a trapped system [Fig. \ref{fig4}(a) and (c)]
and the corresponding {\it ansatz} of Eq. (\ref{kagaome}) [Fig. \ref{fig4}(b) and (d)].
For each spin component, density distributions
show kagome lattice structures.

\section{Fragmented ground states} 

Although we have discussed within the mean-field theory so far,
there is a possibility that a fragmented ground state \cite{nozieres,law1998,ho2000,mueller2006}, 
rather than a mean-field state, arises.
Actually, for the case of SO-coupled spin-1/2 system with an SU(2) symmetry, 
we construct a fragmented state
whose energy is degenerate with the mean-field solution
up to the mean-field approximation.
Such a fragmented ground state is expected to arise in a mesoscopic system.
In this section, we discuss fragmented ground states from the point of view of the time-reversal symmetry.

As pointed out in the Sec.\ref{symmetry}, mean-field states
always break time reversal for the $F=1/2$ case because $\mathcal{T}^2=-1$.
In this section, we show that there are also 
time-reversal invariant many-body states that are
fragmented in a time-reversal preserving $F=1/2$ bosonic system
with or without SOCs.

For a pseudo spin-1/2 system with $N$ atoms, we have $\mathcal{T}^2=(-)^N$.
When $N$ is odd, all states break time-reversal symmetry.
In contrast, when $N$ is even, there is always a time-reversal
invariant ground state which is fragmented:
if $[\mathcal{T},\mathcal{H}]=0$ and $\mathcal{H}|\psi\rangle=E|\psi\rangle$,
$|\psi_{\mathcal{T}}\rangle=|\psi\rangle+\mathcal{T}|\psi\rangle$ is an eigenstate of $\mathcal{H}$ with the same eigenenergy $E$,
and is invariant under time reversal if $N$ is even.
To check whether $|\psi_{\mathcal{T}}\rangle$ is fragmented or not,
we can diagonalize its single-particle density matrix $\hat{\rho}$ \cite{mueller2006}.
We can regroup the single-particle eigenstates as $\{\Psi_{\mu_i},\Psi_{\nu_i}\}$,
where we use $\mu$ and $\nu$ to distinguish two states by their time reversal as
$\mathcal{T}\Psi_{\mu_i}=\Psi_{\nu_i}$ and $\mathcal{T}\Psi_{\nu_i}=-\Psi_{\mu_i}$.
Time-reversal properties of the corresponding creation operators
$\{\hat{a}^{\dag}_{\mu_i},\hat{a}^{\dag}_{\nu_i}\}$ and annihilation operators
$\{\hat{a}_{\mu_i},\hat{a}_{\nu_i}\}$ for the states $\{\Psi_{\mu_i},\Psi_{\nu_i}\}$ are
\begin{eqnarray}
  \mathcal{T}\hat{a}^{\dag}_{\mu_i}\mathcal{T}^{-1}=\hat{a}^{\dag}_{\nu_i},
  \quad
  \mathcal{T}\hat{a}^{\dag}_{\nu_i}\mathcal{T}^{-1}=-\hat{a}^{\dag}_{\mu_i},
  \nonumber\\
  \mathcal{T}\hat{a}_{\mu_i}\mathcal{T}^{-1}=\hat{a}_{\nu_i},
  \quad
  \mathcal{T}\hat{a}_{\nu_i}\mathcal{T}^{-1}=-\hat{a}_{\mu_i}.
  \label{timereversal}
\end{eqnarray}
We also have
$\langle \psi_{\mathcal{T}}|\hat{a}_{\mu_i}^{\dag}\hat{a}_{\mu_j}|\psi_{\mathcal{T}}\rangle
=\langle \mathcal{T}\psi_{\mathcal{T}}|\mathcal{T}\hat{a}_{\mu_i}^{\dag}\hat{a}_{\mu_j}|\psi_{\mathcal{T}}\rangle^*
=\langle \psi_{\mathcal{T}}|\hat{a}_{\nu_i}^{\dag}\hat{a}_{\nu_j}|\psi_{\mathcal{T}}\rangle^*$,
where for the first equality we use the fact that $\mathcal{T}$ is an antiunitary operator,
while the second equality is due to Eq. (\ref{timereversal}) and the fact that
$|\psi_{\mathcal{T}}\rangle$ is invariant under time reversal.
Similarly, we obtain
$\langle \psi_{\mathcal{T}}|\hat{a}_{\nu_i}^{\dag}\hat{a}_{\nu_j}|\psi_{\mathcal{T}}\rangle=
\langle \psi_{\mathcal{T}}|\hat{a}_{\mu_i}^{\dag}\hat{a}_{\mu_j}|\psi_{\mathcal{T}}\rangle^*$,
$\langle \psi_{\mathcal{T}}|\hat{a}_{\mu_i}^{\dag}\hat{a}_{\nu_j}|\psi_{\mathcal{T}}\rangle=
-\langle \psi_{\mathcal{T}}|\hat{a}_{\nu_i}^{\dag}\hat{a}_{\mu_j}|\psi_{\mathcal{T}}\rangle^*$,
and
$\langle \psi_{\mathcal{T}}|\hat{a}_{\nu_i}^{\dag}\hat{a}_{\mu_j}|\psi_{\mathcal{T}}\rangle=
-\langle \psi_{\mathcal{T}}|\hat{a}_{\mu_i}^{\dag}\hat{a}_{\nu_j}|\psi_{\mathcal{T}}\rangle^*$.
These equalities imply that the single-particle density matrix is invariant under $\Xi$ as
\begin{eqnarray}
  \Xi\hat{\rho}\Xi^{-1}=\hat{\rho},\quad \Xi\equiv\mathcal{K}\prod_i\otimes
  \left(
  \begin{array}{cc}
    0 & 1\\
    -1 & 0\\
  \end{array}
  \right).
  \label{densitymatrix}
\end{eqnarray}
Due to this symmetry, we can infer that if $|\Psi\rangle$ is an eigenstate
of $\hat{\rho}$, $\Xi|\Psi\rangle$ is also an eigenstate of $\hat{\rho}$ with
the same eigenenergy but orthorgonal to $|\Psi\rangle$.
Our argument above is also valid for the ground state.

A simple state that preserves time-reversal symmetry is
\begin{eqnarray}
  |\psi_{\mathcal{T}}\rangle\propto \left(\hat{a}_{\mu_1}^{\dag}\right)^{N/2}\left(\hat{a}_{\nu_1}^{\dag}\right)^{N/2}|\rm vac\rangle.
  \label{twoorbital}
\end{eqnarray}
Using this {\it ansatz} wave function, we numerically minimize the energy functional
$\langle\psi_{\mathcal{T}}|\mathcal{H}|\psi_{\mathcal{T}}\rangle$,
and find that the ground-state energy is the same as that obtained with single-orbital
mean-field approximation for an SU(2)-symmetric system without spin-dependent interactions,
where we take the lowest-energy band approximation by considering only states $\Psi_{0,l}$
with $v'=15$, $\beta'=0$, and $\alpha'$ ranging from 0 to 1.

Before conclusions, we would like to point out that the fragmented ground states
we discuss here are protected by time-reversal symemtry for a spin-1/2 system.
The two-orbital ground states for an SU(2)-symmetric spin-1/2 system we constructed in Eq. (\ref{twoorbital})
are different from that predicted in Refs. \cite{kuklov2002,ashhab2003}
in which the fragmented ground states are produced in a cooling process that conserves the total spin $S$.
For an initial state at high temperature, we have $S\sim\sqrt{N}$.
If all atoms condense into the same orbital state, $S=N/2$.
Therefore, to be compatible with the requirement of the total spin conservation,
atoms should condense at least into two orbitals, resulting in
a fragmented ground state.

\section{Summary} 
We have systematically classified ground states of
strong SO-coupled trapped spinor BECs,
including pseudo spin-1/2, spin-1 and spin-2 cases, based
on symmetry analysis. In accordance with breaking of simultaneous SO(2) spin-space rotation symmetry
in favor of discrete symmetries,
there emerge lattice phases showing stripe, triangular,
square, and kagome lattice structures on each spin component.
Imposing symmetries of $\mathcal{T}$, $\mathcal{C}_{nz}$ or $\mathcal{TC}_{nz}$
or combined them with the U(1) gauge transformation
on the order parameter, we predict several lattice phases, some of which are
found to be ground states of a trapped system.
For the spin-1/2 case, the ground states can be classified into
two classes: one breaks time-reversal symmetry
and the other is invariant under time reversal and shows
fragmented properties.

\section{Acknowledgement}
Z.F.X. acknowledges Shunsuke Furukawa for useful discussions,
Shohei Watabe and Nguyen Thanh Phuc for reading the manuscript.
This work was supported by KAKENHI 22340114 and 22740265, a Grant-in-Aid for Scientific Research on Innovation
Areas ``Topological Quantum Phenomena'' (KAKENHI 22103005),
a Global COE Program ``the Physical Sciences Frontier'',
the Photon Frontier Network Program, from MEXT of Japan,
NSFC (No.~91121005 and No.~11004116), and the
research program 2010THZO of Tsinghua University.
Z.F.X. acknowledges the support from JSPS (Grant No. 2301327).
Y.K. acknowledges the support from  Inoue Foundation for Science.


\begin{thebibliography}{10}
\bibitem{bloch2008}
  I. Bloch, J. Dalibard, and W. Zwerger, Rev. Mod. Phys.
  \textbf{80}, 885 (2008).

\bibitem{lin2011}
  Y.-J. Lin, K. Jim\'enez-Garc\'ia, and I. B. Spielman, Nature (London) \textbf{471}, 83 (2011).

\bibitem{dalibard2011}
  J. Dalibard et al., Rev. Mod. Phys. \textbf{83}, 1523 (2011).

\bibitem{zutic2004}
  I. \v{Z}uti\'c, J. Fabian, S. Das Sarma, Rev. Mod. Phys. \textbf{76}, 323 (2004).

\bibitem{qi2010}
  X.-L. Qi and S.-C. Zhang, Physics Today \textbf{63}, 33 (2010).

\bibitem{stanescu2008}
  T. D. Stanescu, B. Anderson, and V. Galitski, Phys. Rev. A
  \textbf{78}, 023616 (2008).

\bibitem{wu2011}
  C.-J. Wu, I. Mondragon-Shem, X.-F. Zhou, Chinese Physics Letters \textbf{28}, 097102 (2011).

\bibitem{wang2010}
  C. Wang {\it et al.}, Phys. Rev. Lett.  \textbf{105}, 160403 (2010).

\bibitem{ho2011}
  T.-L. Ho and S. Zhang, Phys. Rev. Lett. \textbf{107}, 150403 (2011).

\bibitem{yip2011}
  S.-K. Yip, Phys. Rev. A \textbf{83}, 043616 (2011).

\bibitem{xu2011}
  Z. F. Xu, R. L\"u, and L. You, Phys. Rev. A \textbf{83}, 053602 (2011).

\bibitem{kawakami2011}
  T. Kawakami, T. Mizushima, and K. Machida,
  Phys. Rev. A \textbf{84}, 011607(R) (2011).

\bibitem{xqxu2011}
  X.-Q. Xu and J. H. Han, Phys. Rev. Lett. \textbf{107}, 200401 (2011).

\bibitem{hu2012}
  H. Hu {\it et al.}, Phys. Rev. Lett. \textbf{108}, 010402 (2012).

\bibitem{sinha2011}
  S. Sinha, R. Nath, and L. Santos, Phys. Rev. Lett. \textbf{107}, 270401
  (2011).

\bibitem{radic2011}
  J. Radi\'c {\it et al.}, Phys. Rev. A \textbf{84}, 063604 (2011).

\bibitem{zhou2011}
  X.-F. Zhou, J. Zhou, and Congjun Wu, Phys. Rev. A
  \textbf{84}, 063624 (2011).

\bibitem{deng2011}
  Y. Deng {\it et al.}, Phys. Rev. Lett. \textbf{108}, 125301 (2012).

\bibitem{ruseckas2005}
  J. Ruseckas {\it et al.}, Phys. Rev. Lett. \textbf{95}, 010404 (2005);
  G. Juzeli\=unas, J. Ruseckas, and J. Dalibard,
  Phys. Rev. A \textbf{81}, 053403 (2010).

\bibitem{campbell2011}
  D. L. Campbell, G. Juzeli\=unas, and I. B. Spielman, Phys. Rev. A \textbf{84}, 025602 (2011).

\bibitem{sau2011}
  J. D. Sau {\it et al.}, Phys. Rev. B \textbf{83}, 140510(R) (2011).

\bibitem{xu2011b}
  Z. F. Xu and L. You, Phys. Rev. A \textbf{85}, 043605 (2012).

\bibitem{ueda2010}
  Y. Kawaguchi and M. Ueda, e-print arXiv:1001.2072.

\bibitem{volovik1985}
  G. E. Volovik and L. P. Gor'kov, Sov. Phys. JETP \textbf{61}, 843 (1985).

\bibitem{bruder1986}
  C. Bruder and D. Vollhardt, Phys. Rev. B \textbf{34}, 131 (1986). 

\bibitem{makela2007}
  H. M\"akel\"a and K.-A. Suominen, Phys. Rev. Lett. \textbf{99}, 190408 (2007)

\bibitem{yip2007}
  S.-K. Yip, Phys. Rev. A \textbf{75}, 023625 (2007).

\bibitem{kawaguchi2011}
  Y. Kawaguchi and M. Ueda, Phys. Rev. A \textbf{84}, 053616 (2011).

\bibitem{jacob2009}
  A. Jacob, Non-Abelian Atom Optics (Dissertation, Leibnitz, Universit\"at Hannover, 2009).

\bibitem{numerical}
  To obtain mean-field ground states, we take two different
  numerical procedures: (1) using the method of imaginary-time propagation of
  coupled Gross-Pitaevskii equations.  
  For the strong SOC case, we use the fast Fourier transformation
  to calculate the kinetic term more accurately;
  (2) using the simulated annealing method to find the global 
  minimum of the energy functional for the case of strong SOC,
  by taking into account about 30 lowest bases from the Eq. (\ref{eigenstates}).

\bibitem{fractional}
  In spinor condensates, we usually focus on a special phase, such as the polar phase or
  the cyclic phase. Therefore, by simply calculating the total gauge transformation along
  a loop around the vortex core, if the value is $2\pi/\mathbb{N}$ ($\mathbb{N}\in\mathbb{Z}$)
  we know that the vortex here is integer ($\mathbb{N}=1$) or fractional ($\mathbb{N}>1$).
  Due to the spin-orbit coupling, different phases, such as ferromagnetic, polar, and etc,
  appear in the same ground state, which makes previous arguments on spinor condensates
  invalid. 

\bibitem{statement}
  To check whether the state $\psi$ of Eq. (\ref{orderparameters}) is invariant under a specific operator
  $\hat{O}$ or not, we need to compare $\psi$ and $\hat{O}\psi$.
  If $\psi=\hat{O}\psi$, we infer that $\psi$ is invariant under $\hat{O}$.
  Due to the noncommutative property of the U(1) gauge transformation and time reversal,
  the global U(1) gauge transformation will not change the structure of the ground 
  states, but can change the combined symemtry of U(1) gauge transformation and time reversal,
  as if we have $\mathcal{T}\psi=\psi$, by doing a global U(1) gauge transformation $\psi\rightarrow\exp(i\phi)\psi$,
  and now $\exp(2i\phi)\mathcal{T}\exp(i\phi)\psi=\exp(i\phi)\psi$.
  Therefore we neglect the U(1) phase when discussing time-reversal symmetry or the combined time-reversal symmetry
  and spin-space rotation symmetry.

\bibitem{nozieres}
  P. Nozi\`eres and D. Saint James, J. Phys. France \textbf{43}, 1133-1148 (1982);
  P. Nozi\`eres, {\it Bose-Einstein Condensation}, edited by A. Griffin, D. W. Snoke, and S. Stringari,
  (Cambridge University Press, New York, 1996).


\bibitem{law1998}
  C. K. Law, H. Pu and N. P. Bigelow, Phys. Rev. Lett. \textbf{81}, 5257 (1998).

\bibitem{ho2000}
  T. L. Ho and S. K. Yip, Phys. Rev. Lett. \textbf{84}, 4031 (2000).
  
\bibitem{mueller2006}  
  E. J. Mueller, {\it et al.}, Phys. Rev. A \textbf{74}, 033612 (2006).

\bibitem{kuklov2002}
  A. B. Kuklov and B. V. Svistunov, Phys. Rev. Lett. \textbf{89}, 170403 (2002).

\bibitem{ashhab2003}
  S. Ashhab and A. J. Leggett, Phys. Rev. A \textbf{68}, 063612 (2003). 

\end{thebibliography}
\end{document}